\documentclass[11pt]{amsart} 
\usepackage{amsmath,amssymb,amsthm}
\usepackage{pdfsync}
\usepackage{enumerate}
\usepackage{graphicx}
\usepackage{latexsym}
\usepackage{amsfonts}
\usepackage[utf8]{inputenc}
\usepackage{amsthm}
\usepackage{amsmath}
\usepackage{amssymb}
\usepackage{xcolor}
\usepackage[
            unicode,
            breaklinks=true,
            colorlinks=true]{hyperref}
\usepackage{amscd}
\usepackage[all]{xy} % commutative diagrams
\usepackage{mathrsfs} % calligraphic capitals
\usepackage{stmaryrd} % extra symbols
\usepackage{amsopn} % define new math operators
\usepackage{eepic}
\usepackage{epic}
\usepackage{eucal}
\usepackage{epsfig}
\usepackage{psfrag}

\newtheorem{theorem}{Theorem}[section]

\newtheorem{lemma}[theorem]{Lemma}

\newcommand{\R}{{\mathbb R}}

\newcommand{\T}{{\mathbb T}}

\newcommand{\Hessien}{{\operatorname{Hess}}}

\newcommand{\sign}{{\operatorname{sign}}}

\newcommand{\ham}[1]{\mathcal{X}_{#1}}

\newcommand{\op}[1]{\!\!\mathop{\rm ~#1}\nolimits}
\newcommand{\DD}{\mathrm{d}}

\renewcommand{\leq}{\leqslant}

\newenvironment{remark}{\refstepcounter{theorem}\par\medskip\noindent{\bf
Remark~\thetheorem.}}{\unskip\nobreak\hfill\hbox{}}

\newenvironment{question}{\refstepcounter{theorem}\par\medskip\noindent{\bf
Question~\thetheorem~~}}{\unskip\nobreak\hfill\par}

\newenvironment{definition}{\refstepcounter{theorem}\par\medskip\noindent{\bf
Definition~\thetheorem.}}{\unskip\nobreak\hfill\hbox{}}
\begin{document}

\title[Generating hyperbolic singularities]{Generating hyperbolic singularities in completely integrable systems}

 \author{Holger R.~Dullin\,\,\,\,\,\,\,\,\,\,\,\,\,\,\,\,\,\,\,\'Alvaro Pelayo} \date{}

\maketitle
\thispagestyle{empty}

\begin{abstract}
Let $(M,\Omega)$ be a connected symplectic $4$\--manifold and let $F=(J,H) \colon M \to
\mathbb{R}^2$ be a completely integrable system on $M$ with only non\--degenerate
singularities and for which $J \colon M \to \mathbb{R}$ is a proper map. Assume that $F$
does not have singularities with hyperbolic blocks and that  $p_1,\ldots,p_n$ are 
 the focus\--focus singularities of $F$. For each subset $S=\{i_1,\ldots,i_j\}$ we will show how to modify $F$ locally
around any $p_i, i \in S$, in order to create a new integrable system $\widetilde{F}=(J, \widetilde{H}) 
\colon M \to \mathbb{R}^2$ such that its classical spectrum $\widetilde{F}(M)$  contains $j$ smooth curves
of singular values corresponding to non\--degenerate transversally hyperbolic singularities of $\widetilde{F}$.
Moreover the focus\--focus singularities of $\widetilde{F}$ are precisely $p_i$, $i \in \{1,\ldots,n\} \setminus S$, and each of these $p_i$ is non\--degenerate. The proof is based on Eliasson's
linearization theorem for non\--degenerate singularities, and properties of the Hamiltonian Hopf bifurcation.
 \end{abstract}

\section{Introduction}
 
 This paper intends to shed some light on the following question in the theory of
 finite dimensional completely integrable Hamiltonian systems:
 
 \begin{question}
\emph{Suppose that $f_1,\ldots,f_n \colon M \to \mathbb{R}$ form an integrable system on a $2n$\--dimensional symplectic manifold $(M,\omega)$. What is the relation between the following conditions?}
 \begin{itemize}
  \item[(1)]
\emph{the flow of at least one of the functions $f_i$, $i \in \{1,\ldots,n\}$, is periodic;}
\item[(2)]
\emph{existence of a set $B \subset M$ of non\--degenerate focus\--focus singularities of the joint map $(f_1,\ldots,f_n) \colon M \to \mathbb{R}^n$;}
 \item[(3)]
 \emph{existence of a set $C \subset M$ of non\--degenerate hyperbolic singularities of $(f_1,\ldots,f_n)$;}
 \item[(4)]
 \emph{existence of set $D \subset M$ of degenerate singularities of $(f_1,\ldots,f_n)$.}
 \end{itemize} 
 \end{question}
 Several results are known which give partial answers to this question. For instance,  under general assumptions on $f_1,\ldots,f_n$ and  $M$,  if condition (1) 
 holds for all $i=1,\ldots,n$,  then $$B=C=D=\varnothing.$$ In fact, there is a complete theory for these
 systems - called \emph{toric integrable systems} - when $M$ is compact, by Atiyah, Guillemin-Sternberg, and Delzant \cite{At82,GuSt82, De88}. 
 
 The results of this paper imply that there are 
 integrable systems for which (1), (2), (3), and (4) hold simultaneously with $B\neq \varnothing$, $C\neq \varnothing$,
 and $D \neq \varnothing$
 (this is the content of Theorems~\ref{gen:thm}, \ref{mt2}, and \ref{thm:main}). Moreover, it seems quite plausible that
 under some general condition, (1), (2) and (3) with $B\neq \varnothing$ and $C\neq \varnothing$ will imply that $D \neq \varnothing$ in (4) (see Question~\ref{hc}).

 Because several of the technical tools we use are exclusive to dimension four, from now
 on we assume that $2n=4$. Let $(M, \Omega)$ be a  connected symplectic $4$\--manifold, that is, $M$ is a smooth connected
$4$\--manifold, and $\Omega$ is a non\--degenerate closed $2$\--form on $M$, i.e. a \emph{symplectic
form}.  A smooth function $f \colon M \to \mathbb{R}$ induces a vector field $\ham{f}$ on $M$
by means of Hamilton's equation:
$$
\Omega(\ham{f},\, \cdot)=-{\rm d}\!f.
$$
 The vector field $\ham{f}$ is called
the \emph{Hamiltonian vector field} induced by $f$.   Given any two smooth functions
$f,g \colon M \to \mathbb{R}$ we may define their \emph{Poisson bracket}
$$\{f,\,g\}:=\Omega(\ham{f},\, \ham{g}).$$

An
\emph{integrable system}\footnote{More precisely, a ``finite dimensional completely integrable Hamiltonian system".} is a triple $$(M,\Omega,(J,H))$$  where $(M,\Omega)$ is
a connected symplectic $4$\--manifold and $J,H \colon M \to \mathbb{R}$ are smooth
functions  for which
$\{J,\,H\}\equiv 0$  on $M$, and such that
the differentials ${\rm d}J$, ${\rm d} H$ are 
linearly independent almost everywhere on $M$.  
Near each point in $M$ there are coordinates $(x,\, y,\, \xi,\, \eta)$ in which
the symplectic form $\Omega$ is given by ${\rm d} \xi \wedge
{\rm d}x +{\rm d} \eta\wedge {\rm d}y$, and the equation $\{J,\,H\}\equiv 0$ 
may be written as  a partial differential
equation
  $$
  \frac{\partial J}{\partial \xi} \, \frac{\partial H}{\partial x} -
  \frac{\partial J}{\partial x} \, \frac{\partial H}{\partial \xi} +
  \frac{\partial J}{\partial \eta} \, \frac{\partial H}{\partial y} -
  \frac{\partial J}{\partial y} \, \frac{\partial H}{\partial \eta}
  =0,
  $$
  which is equivalent to $J$ (respectively $H$) being
  constant along the flow lines of $\ham{H}$ (respectively $\ham{J}$).  The following
  result gives a method to attach hyperbolic singularities to an integrable
  system, by modifying locally the system near its focus\--focus singularities. Focus\--focus
  singularities come endowed with a Hamiltonian circle action near the focus\--focus
  singular fiber, it is the Hamiltonian action generated by the $J$\--component 
  of the system in local normal form (as explained in Section~\ref{Els}).  The image
  of the joint map $F:=(J,M) \colon M \to \mathbb{R}^2$ is called by physicists
  \emph{the classical spectrum} of $F$ (in analogy with the semiclassical
  spectrum of quantum mechanics).

  Throughout
  the paper we assume that each focus\--focus singularity is in a singular fiber
  in which it is the only singularity (a system satisfying this property is often called
  ``simple"). Topologically, this means that the singular fiber containing a focus\--focus
  singularity is a torus pinched precisely once (as opposed to a multipinched torus).

\begin{theorem} \label{gen:thm}
Let $(M,\Omega, F:=(J,H) \colon M \to
\mathbb{R}^2)$ be an integrable system on a connected symplectic $4$\--manifold 
$(M,\Omega)$ with only non\--degenerate
singularities and for which $J \colon M \to \mathbb{R}$ is a proper map. Assume that $F$
does not have singularities with hyperbolic blocks and that $p_1,\ldots,p_n$ are all the focus\--focus
singularities of $F$, each of which is simple. For each subset $S=\{i_1,\ldots,i_j\}$, $1 \leq j \leq n$, there is an integrable system 
$$(M,\Omega,\widetilde{F}_S:=(J, \widetilde{H}) 
\colon M \to \mathbb{R})$$  such that its  classical spectrum $\widetilde{F}_S(M)$  contains $j$ loops consisting of three piecewise smooth 
curves of singular values. One smooth piece corresponds to non\--degenerate transversally hyperbolic singularities of $\widetilde{F}_S$
and two pieces correspond to non\--degenerate transversally elliptic singularities.
Moreover the focus\--focus singularities of $\widetilde{F}_S$ are precisely $p_i$, $i \in \{1,\ldots,n\} \setminus S$, and each $p_i, i \in  \{1,\ldots,n\} \setminus S$, is non\--degenerate.
\end{theorem}
  
  The proof of Theorem~\ref{gen:thm} uses Eliasson's
linearization theorem for non\--degenerate singularities, and properties of the Hamiltonian Hopf bifurcation.
  
  \begin{remark}
   The
  properness of $J$ in Theorem~\ref{gen:thm} means that the preimage under $J$ of a compact set is
  compact in $M$. 
  \end{remark}

  \medskip
  
  \emph{Structure of the paper}.
   In Section \ref{sing:sec} we review the different types of singularities
  an integrable system can have, state Eliasson's linearization theorem, and recall the notion of degenerate singularity.
In Section~\ref{hyp} we introduce 
  hyperbolic semitoric systems, and prove a version of Theorem~\ref{gen:thm} for semitoric systems
  that takes into account the additional symmetry given by the semitoric property.     
  In Section~\ref{Els} we explain the existence of a global Hamiltonian circle action semiglobally near
  a singularity of focus\--focus type, using Eliasson's linearization theorem.
  In Section~\ref{mt:sec} we state
  the general theorem of the paper, Theorem~\ref{thm:main}, of which Theorem~\ref{gen:thm} is a consequence. In Section~\ref{sec:Hopf}
we briefly recall the basics about Hopf bifurcations which we need for the proof of Theorem~\ref{thm:main}. The remaining
of the paper is devoted to the proof of Theorem~\ref{thm:main}.
   
   \medskip
   
  \emph{Acknowledgements}. 
  We would like to thank Joachim Worthington for his contribution in the analysis of 
  the example in section~\ref{sec:Examples}.
  The first author is partially supported by ARC grant DP110102001.
  The  second author is partially supported by NSF grants DMS-1055897 and DMS-1518420.

  \section{Eliasson's Linearization of singularities} \label{sing:sec}
  
 Let $(M,\Omega,F:=(J,H) \colon M \to \mathbb{R}^2)$ be an integrable system and
 let $p\in M$.  We say that $p$ is a \emph{regular point} if ${\rm d}_pF$ has
rank $2$.  A point $c \in \mathbb{R}^2$ is called a \emph{regular value} if 
every point in $F^{-1}(c)$ is regular; in this case,  $F^{-1}(c)$ is called a \emph{regular fiber}.  We say that 
$p$ is a \emph{critical point} (or a \emph{singularity}), if the rank of 
$\op{d}_pF$ is $0$ or $1$. The fiber $F^{-1}(c)$ is a \emph{singular fiber} if it contains 
one or more critical points

Let $X$ be a connected component of a regular fiber $F^{-1}(c)$ and assume that 
the Hamiltonian vector fields $\mathcal{X}_{f_1}, \mathcal{X}_{f_2}$
are complete on $F^{-1}(c)$. Then it follows from the definition of integrable
system that $X$ is diffeomorphic to $\T^2$, $S^1 \times \R$, or $\R^2$. 
In the particular case that ${F}^{-1}(c)$
is compact, then $\mathcal{X}_{f_1}, \mathcal{X}_{f_2}$ are
complete, and therefore $X$ is diffeomorphic to
$\mathbb{T}^2$; this is always true for instance if at least one of $f_1, f_2, F$ is a proper map.

\subsection{Non degeneracy}

Throughout this paper we use the notion of \emph{non\--degeneracy} of a singularity. 
Suppose that $(M, \Omega)$ is a connected symplectic 
four-manifold.  Let $F:=(f_1,f_2)$ be an integrable system on 
$(M, \Omega)$ and $p \in M $ a critical point of $F$.

If 
$\DD_pF=0$, then $p$ is called \emph{non-degenerate} if
the Hessians $\operatorname{Hess}f_j(p)$, $j=1,2$, span a Cartan 
subalgebra of the symplectic Lie algebra of quadratic forms 
on the tangent space $({\rm T}_p M, \Omega_p)$ to $M$ at $p$. 
It follows from the work of Williamson~\cite{williamson}  that  such a Cartan
subalgebra 
has a basic building block of three types: two
uni-dimensional ones (the elliptic block $x^2+\xi^2$ and the real
hyperbolic block: $x\xi$) and a two-dimensional block called
focus-focus: 
\begin{align*}
J_1&=x\eta-y\xi\\
J_2&=x\xi+y\eta.
\end{align*}

If $\operatorname{rank}\left(\DD_pF\right)=1$ one may assume 
that $\DD_p f_1 \neq0$.  Let $\iota \colon S \to M$ be an 
embedded local $2$\--dimensional symplectic submanifold through 
the point $p$, such that ${\rm T}_p S\subset \ker(\DD_p f_1)$ and 
${\rm T}_pS$ is transversal to the Hamiltonian vector field
$\mathcal{H}_{f_1}$ defined by $f_1$. The 
critical point $p$ of $F$ is transversally 
\emph{non-degenerate} if $\operatorname{Hess}(\iota^\ast f_2)(p)$ 
is a non-degenerate symmetric bilinear form on ${\rm T}_p S$. 
 
\begin{remark}
The notion of non\--degeneracy is independent of the choice we make of $S$. 
The existence of $S$ follows from the Flow Box Theorem 
(see, eg. \cite[Theorem 5.2.19]{AbMa1978}, also called  
the Darboux-Carath\'eodory theorem \cite[Theorem 4.1]{PeVN11}).
\end{remark}

\begin{remark}
 For the notion of non-degeneracy of a critical point 
in arbitrary dimension see \cite{vey}, 
\cite[Section 3]{san-mono}. 
\end{remark}

\subsection{Linearization theorem}

% ---------------------------------------
\begin{figure}[h]
  \centering
  \includegraphics[width=9cm]{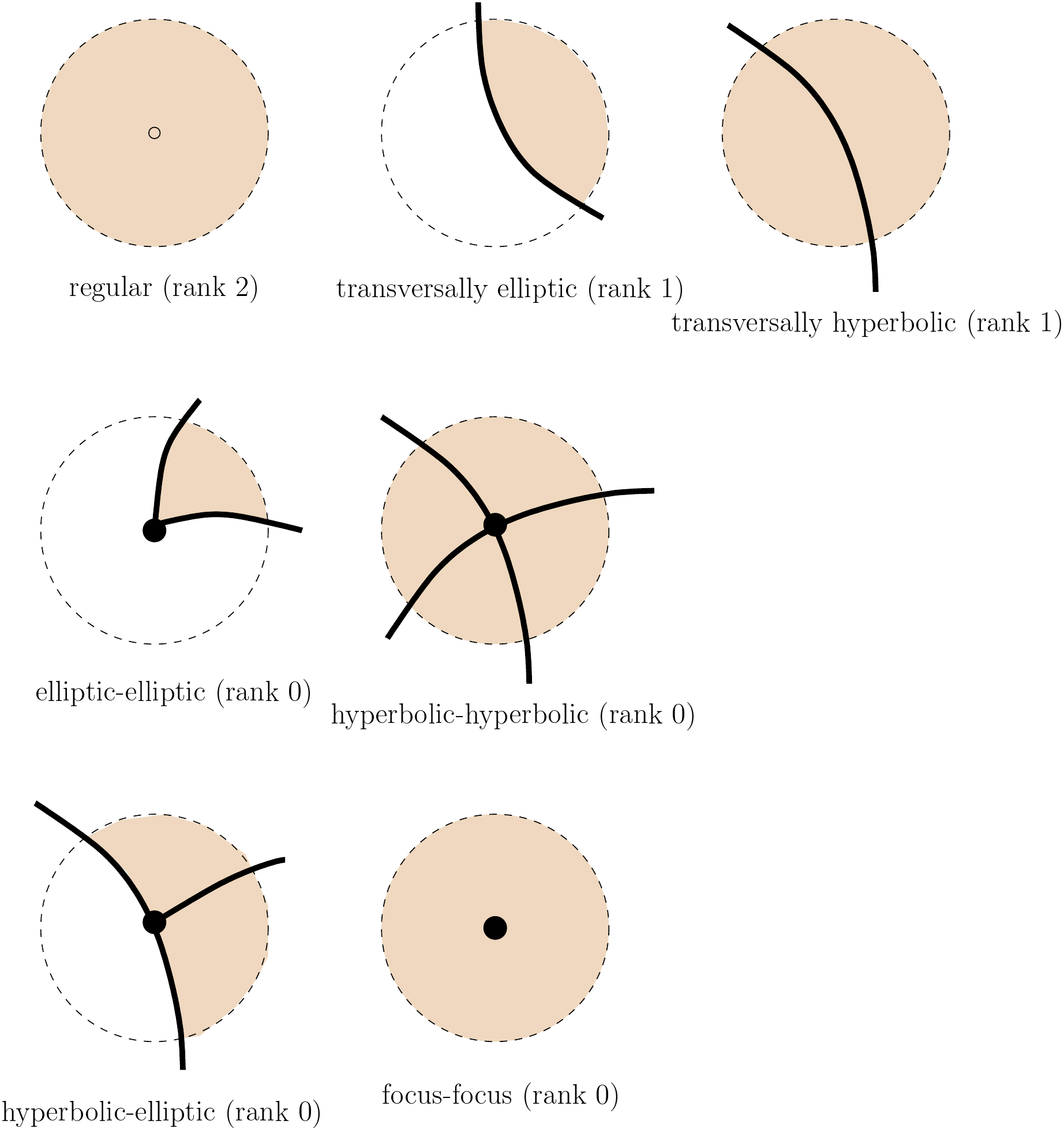}
  \caption{Local description of the possible images of $F$ near a regular or singular value.
   The transversally\--hyperbolic, elliptic\--hyperbolic, and
    hyperbolic\--hyperbolic cases are not possible if $F$ is semitoric
    (Definition~\ref{d}), but they are possible if $F$ is hyperbolic
    semitoric (Definition~\ref{def:ANDSISH}).}
  \label{singularities_in_image}
\end{figure}
% ---------------------------------------

%
Non-degenerate critical points may be characterized (\cite{El90, El84, VNWa14}) 
using the Williamson normal form \cite{williamson}.  
Eliasson's theorem stated below describes the local normal form
around non\--degenerate singularities of an integrable system; the result
holds in any dimension, but we state it here in dimension four since it
is the case we are concerned with in the present paper.
The analytic version of the theorem is due to Vey~\cite{vey}.

\begin{theorem}[Eliasson \cite{El90,El84, VNWa14}] \label{Eliasson}
  The non\--degenerate critical points of an integrable
  system $F \colon M \to \mathbb{R}^2$ are linearizable. That is, if $p
  \in M$ is a non\-degenerate critical point of the 
  integrable system $F=(f_1,f_2): M \rightarrow \mathbb{R}^2$
  then there exist symplectic coordinates $(x,y,
  \xi, \eta)$ near $p$, in which $p$ is represented as
  $(0,0,0, 0)$, such that $\{f_i,\,J_j\}=0$, for all
  $i,\,j \in \{1,2\}$, where we have the following possibilities for each $J_i$, $i \in \{1,2\}$, each of which is defined on a 
  neighborhood of $(0,0,0,0) \in \mathbb{R}^4$:
  \begin{itemize}
  \item[{\rm (i)}]  Elliptic component: $J_i = (x^2 + \xi^2)/2$ or $J_i=(y^2+\eta^2)/2$.
  \item[{\rm (ii)}]  Hyperbolic component: $J_i = x \xi$ or $J_i=y\eta$.
    \item[{\rm (iii)}] Focus\--focus component: $J_{1}=x\eta-  y\xi$ and $J_{2} =x\xi+y\eta$     (note that this component appear in ``pairs'').
  \item[{\rm (iv)}] Non\--singular component: $J_{i} = \xi$ or $J_i=\eta$.
  \end{itemize}
  Moreover if $p$ does not have any hyperbolic block, then the system
  of commuting equations $\{f_i,\,J_j\}=0$, for all indices $i,\,j \in \{1,2\}$,
  may be replaced by the single equation
 $$
 (F-F(p))\circ \varphi = g \circ (J_1,\,J_2),
 $$ 
 where $\varphi=(x,y,\eta,\xi)^{-1}$
 and $g$ is a diffeomorphism from a small neighborhood of the origin
 in $\mathbb{R}^4$ into another such neighborhood, such that $$g(0,0,0,0)=(0,0,0,0).$$
\end{theorem}

 See Figure~\ref{singularities_in_image} for a local description of the singularities
 appearing in Theorem~\ref{Eliasson}.

\medskip

 \begin{remark}
The analytic case of Theorem~\ref{Eliasson} was proved by
R{\"u}{\ss}mann in \cite{russmann} when $2n=4$, and then in any dimension by Vey~\cite{vey}. 
 \end{remark}

 A simple way to check Eliasson non-degeneracy is as follows:
 
 \begin{lemma} \label{lem:xxx}
 Let $F:=(f_1,f_2) \colon M \to \mathbb{R}^2$ be an integrable system.
A critical point $p$ of $F$
of rank $0$ is non\--degenerate if the Hessians $\Hessien f_1(p)$ 
and $\Hessien f_2(p)$ are linearly independent, 
and there is a linear combination 
 $$\alpha B \, \Hessien f_1(p) + \beta B \, \Hessien f_2(p)$$
 for which there are no multiple eigenvalues.
 Here $B$ is the symplectic matrix corresponding to the symplectic form $\Omega$.
 In particular if $B \Hessien f_1(p)$ has no multiple eigenvalues then 
 $p$ is non-degenerate.
 \end{lemma}
 
 This criterion is based on the fact (see for instance \cite{BolsinovFomenko04})
 that a commutative subalgebra 
 of the symplectic algebra is a Cartan subalgebra if and only if it is 
 two-dimensional and if it contains an elements whose eigenvalues are different.

  \section{Hyperbolic semitoric systems} \label{hyp}

Theorem~\ref{gen:thm} may be applied in particular to enlarge the category of \emph{semitoric systems}, these
are systems which have an additional circular Hamiltonian symmetry coming from a global Hamiltonian
action of the circle $S^1$. Many integrable systems from classical mechanics (see eg. the book by Holm \cite{Ho11}, or 
the article \cite{PeVN12b}),
exhibit symmetries of this nature including the Lagrange Top, the two\--body problem, and the spherical pendulum.
 Recall that an action of the circle $S^1$ on a symplectic manifold $(M,\Omega)$
by symplectomorphisms is \emph{Hamiltonian} if   there exists a smooth map $J \colon M \to \mathbb{R}$, called
the \emph{momentum map} such that $$\Omega(X_M,\cdot)=-{\rm d}J,$$ where $X_M$ is the vector
field (or infinitesimal generator) of the $S^1$\--action.
The category of semitoric systems includes many examples
from the physics literature such as integrable systems of Jaynes\--Cummings type, and the Jaynes\--Cummings
system.  In this section we state a theorem which allows us to construct a semitoric system with
hyperbolic singularities, from a semitoric system without hyperbolic singularities; below we give the
precise definitions of these notions.

\subsection{The Jaynes\--Cummings system} \label{jc:sec}
The famous \emph{Jaynes\--Cummings system} \cite{JaCu63, Cu65} is given on phase space $S^2 \times \mathbb{R}^2$ by 
\begin{eqnarray} \label{jc}
J := \frac{u^2+v^2}{2} +  z
\,\,\,\,\, \textup{and}\,\, \,\,\,H := \frac{1}{2} \, (ux+vy),
 \end{eqnarray}
where $S^2$ is the unit sphere in $\R^3$ with coordinates $(x,\,y,\,z)$,
and  $\R^2$ is equipped with coordinates $(u,\, v)$. We endow $S^2\times\R^2$ with the product
symplectic structure $\omega_{S^2} \oplus  \omega_0$ where $\omega_{S^2}$ is the standard
symplectic form on $S^2$ and $\omega_0$ is the standard
symplectic form on $\R^2$.  The $J$ component is the momentum map of the Hamiltonian
$S^1$\--action that simultaneously rotates
about the vertical axes of $S^2$ and the origin in
$\mathbb{R}^2$.   Under the flow of $J$, the points $(x,\,y,\,z)$ and
$(u,\,v)$ move along the flows of $z$ and $(u^2+v^2)/2$,
respectively, with same angular velocity, so
$\langle (x,\,y),\, (u,\,v) \rangle=ux+vy=2H$ is constant and
commutes with $J$. The completely integrable system $F \colon S^2 \times \mathbb{R}^2$ given by  (\ref{jc})
has been extensively studied, and recently attracted a lot of interest in both
the physics and mathematics communities, see for instance \cite{BaDo13a, BaDo13b, PeVN12}.
In \cite[Corollary~2.2]{PeVN12} it was proved that $(S^2 \times \R^2, \omega_{S^2} \oplus  \omega_0, (J,H))$
is an example of a so called \emph{semitoric system}, which we define in general next.

\subsection{Semitoric systems}

We start with the notion of a semitoric system (\cite[Definition 2.1]{PeVN09}).

\begin{definition} \label{d}
  A {\em semitoric system} (or \emph{semitoric integrable system}) is an integrable system $(M,\Omega,F:=(J,H) \colon M \to \mathbb{R}^2)$ such
  that:  \begin{itemize}
  \item[{\rm (i)}]
  $J$ is the momentum map for a
  Hamiltonian $S^1$\--action;
  \item[{\rm (ii)}]
  $J$ is proper;
  \item[{\rm (iii)}]
$F$ has only non\--degenerate singularities, without hyperbolic blocks.    
  \end{itemize}
  \end{definition}

  If $F$ is a semitoric system in the sense of
Definition~\ref{d}, we have the following possibilities for the map
$(J_1,\,J_2)$ in Theorem~\ref{Eliasson}, depending on the rank of the singularity:
\begin{itemize}
\item[{\rm (1)}] if $p$ is a singularity of $F$ of rank zero, then
  the building blocks are
  \begin{itemize}
  \item[{\rm (i)}] $J_1 = (x^2 + \xi^2)/2$ and $J_2 = (y^2 +
    \eta^2)/2$.
  \item[{\rm (ii)}] $J_1=x \eta - y \xi$ and $J_2 =x \xi+ y\eta$;  %\,\, on the other hand,
  \end{itemize}
\item[(2)] if $p$ is a singularity of $F$ of rank one, then
  \begin{itemize}
  \item[{\rm (iii)}] $J_1 = (x^2 + \xi^2)/2$ and $J_2 = \eta$.
  \end{itemize}
\end{itemize}

Semitoric systems as in Definition~\ref{d} are classified 
  in \cite{PeVN09,PeVN11} in terms of five symplectic invariants. This classification
  excluded systems with hyperbolic singularities, which are however prominent
  in the physics literature.  By eliminating
  conditions (ii) and (iii) in Definition~\ref{d} one would expand significantly the class of integrable systems 
  which are ``semitoric" in spirit, that is, those for which the component $J$ still generates a
  Hamiltonian $S^1$\--action.   For the purpose of this paper, we keep item (ii) due technical reasons\footnote{it can probably
  be weakened to the condition that $F$ is proper as a map into $\mathbb{R}^2$, as in \cite{PeRaVN14,PeRaVN15}, but this
  still requires substantial work.}, and show that there are many interesting examples of systems which
  satisfy the assumptions (i) and (ii) but not (iii); precisely, we will be concerned with 
  \emph{hyperbolic semitoric systems}.

   \begin{definition} \label{def:ANDSISH}
 A \emph{hyperbolic semitoric system} is an integrable system
$(M, \Omega, F := (J, H) \colon M \to \mathbb{R}^2)$ for which:
\begin{itemize}
\item[{\rm (a)}]
 $J$ is the momentum map for 
a Hamiltonian $S^1$\--action;
\item[{\rm (b)}]
 $J$ is proper;
\item[{\rm (c.1)}]
the set of hyperbolic singularities of $F$ is non\--empty;
\item[{\rm (c.2)}]
the set of degenerate singularities $F$ is isolated.
 \end{itemize}
 \end{definition}

Our next goal is to construct hyperbolic semitoric systems from semitoric systems.

\subsection{From semitoric to hyperbolic semitoric systems}

The main theorem of this paper  (Theorem~\ref{thm:main}) implies that
 any semitoric system in the more strict sense of Definition~\ref{d} may be suitably modified
  to create a new ``semitoric" system in the more relaxed sense of Definition~\ref{def:ANDSISH}.   
  That is, we have the following result.

  \begin{theorem} \label{mt2}
  Given a semitoric system $$(M, \Omega, F:=(J, H))$$ with focus\--focus singularities $p_1,\ldots, p_{m_f}$ and an integer
  $1 \leq k < m_f$, there exists a smooth Hamiltonian function $\widetilde{H}^k \colon M \to \mathbb{R}$
  such that  $$(M, \Omega, \widetilde{F}:=(J, \widetilde{H}^k))$$ is a hyperbolic semitoric system with non\--degenerate
  focus\--focus singularities at $p_1,\ldots,p_k$.
  \end{theorem}
  
  The new aspect of this theorem is that it says that the global Hamiltonian $S^1$\--action of
  the original semitoric system  may be preserved, in addition to the other information about the old and newly created
  singularities given in Theorem~\ref{gen:thm}.

\begin{figure}[htbp]
  \begin{center}
    \includegraphics[width=10cm]{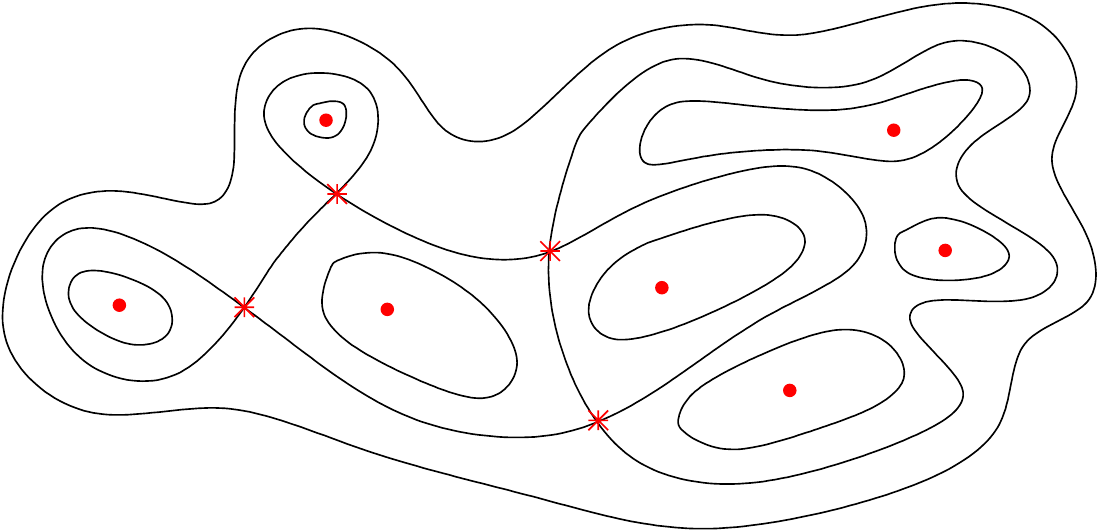}
    \caption{The figure displays elliptic (red dot) and hyperbolic singularities (red star) on a surface.}
    \label{hyperbolic-global}
  \end{center}
\end{figure}

  \section{Focus\--focus singularities and Hamiltonian $S^1$\--actions} \label{Els}
  
  %%%%%%%%%%%%

  The proof in Section~\ref{XX} is based on the knowledge of Eliasson's
  normal form near a  focus\--focus singular point. Here we quickly
  review the ingredients we use.
    
  \begin{definition} \label{mm}
  A smooth map $F = (H_1,H_2) \colon  M \to \mathbb{R}^2$ is a \emph{momentum map} on $U$
  if ${\rm d}F$ is surjective almost everywhere in $U$ and $\{H_1,H_2\}=0$.  
  \end{definition}

  \begin{definition} \label{ll}
A \emph{singular Liouville foliation} $\mathcal{F}$ is a union of connected subsets of $M$, called
  the \emph{leaves} of the foliation, which
  are pairwise disjoint and such that
 there is a momentum map $F\colon W \to \mathbb{R}^2$ in the sense of Definition~\ref{mm} such that the leaves of $\mathcal{F}$ 
 coincide with the connected
components of the fibers $F^{-1}(c)$ for $c$ varying in an open subset of $\mathbb{R}^2$.
 \end{definition}  
  
  \smallskip
    
    Let $(M,\Omega,F:=(f_1,\,f_2))$
  be an integrable system on a connected symplectic $4$\--manifold $(M,\Omega)$ (semitoric
  or not).  
  Let $\mathcal{F}$ be the
  singular Liouville foliation of $M$ associated  to $F$, as in Definition~\ref{ll}.  The leaves of this foliation are the
  connected components of the fibers $F^{-1}(c)$.  Let $p$ be a critical point of focus-focus type.  For 
  simplicity suppose that $F(p)=0$ and that the 
  fiber $\Lambda_0:=F^{-1}(0)$  contains no critical point other than $p$.
  It is well known that  $\Lambda_0$ is a ``pinched torus" (that is,  an immersion of the sphere $S^2$ with a
    transversal double point). All the fibers of $F$ in a neighborhood of the fiber  $\Lambda_0$
    are  $2$\--tori.
     By Theorem~\ref{Eliasson}
  there exist symplectic coordinates $(x, y, \xi,\eta)$ in a
  neighborhood $V$ about the focus\--focus point $p$ in which if
  \begin{align}
    J_1&=x\eta-y\xi \label{equ:cartan1}\\
     J_2&=x\xi+y\eta
    \label{equ:cartan2}
  \end{align}
  then $(J_1,J_2)$ is a momentum map for $\mathcal{F}$; here the critical
  point $p$ corresponds to  $(0,\,0,\,0,\,0)$.    Near $p$, the Hamiltonian flow of $J_1$ is periodic, and assuming V to be
invariant with respect to this flow  the associated $S^1$\--action is free in $V\setminus \{p\}$.
  That is, \emph{we always have  existence of a Hamiltonian action of the circle $S^1$ that commutes
  with the flow of the system semiglobally near a focus\--focus singularity} (i.e.\ in a neighborhood of the singular fiber
  that contains the focus\--focus singular point $p$). 
 This is a property of focus\--focus singularities which is essential for 
 Theorem~\ref{mt2} and our upcoming Theorem~\ref{thm:main}.

  Fix  a point $A\in
  \Lambda_0\cap (V\setminus\{p\})$. Let $\Sigma$ denote a small
  2\--dimensional manifold transversal to $\mathcal{F}$ at 
  $A$.
 Since $\mathcal{F}$ in a neighborhood of $\Sigma$ is
  regular for $F$ and $(J_1,\,J_2)$ simultaneously, there exists 
  a diffeomorphism $\varphi$ from a neighborhood $U$ of $F(A)$ into a
  neighborhood of $(0,0)$ in $\R^2$ such that $(J_1,J_2)=\varphi \circ
  {F}$. Thus there exists a smooth momentum map $\Phi=\varphi
  \circ{F}$ (in the sense of Definition~\ref{mm}) for $\mathcal{F}$, defined on a neighborhood
  $F^{-1}(U)$ of $\Lambda_0$, which agrees with $(J_1,J_2)$ on $V$.
  Write $\Phi:=(H_1,\,H_2)$ and $\Lambda_c:=\Phi^{-1}(c)$ (notice that  $\Lambda_0=\mathcal{F}_p$). 
  It follows from equations~(\ref{equ:cartan1}) and (\ref{equ:cartan2})
  that near $p$ the orbits corresponding to the Hamiltonian $H_1$ are periodic. On the other hand, 
  the vector field $\mathcal{X}_{H_2}$ is hyperbolic and it has a local stable manifold in the $(\xi,\eta)$\--plane
  transversal to its local unstable manifold in the
  $(x,y)$\--plane. Moreover, the vector field $\mathcal{X}_{H_2}$ is radial in the sense
  that the flows approaching the origin do not spiral on
  the local (un)stable manifolds (for further details see \cite[Section 3]{VN03} and
  \cite[Section 5.2.3]{PeVN11}).

%%%%%%%%%%%%

 \section{Main theorem} \label{mt:sec}
 
 \begin{figure}
\begin{center}
\includegraphics[width=8cm]{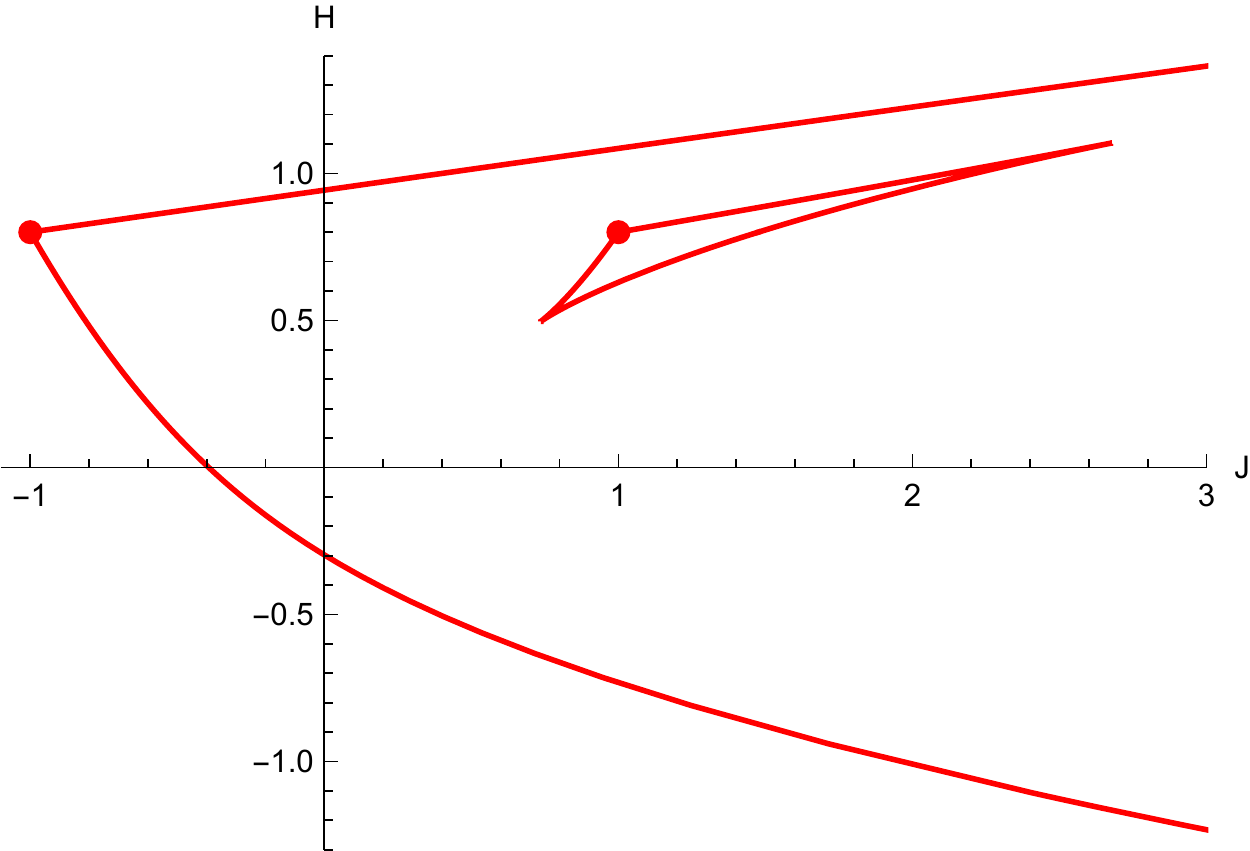}
\end{center}
\caption{The image 
$\hat F(M)$ for an integrable deformation of the spin-oscillator with $G(z) = \frac45 z^2$, illustrating Theorem~\ref{thm:main}.
See Section~\ref{jc:sec} and Section~\ref{sec:Examples}.}
\label{fig:spinoscHopf}
\end{figure}

Note that, by definition, a semitoric system can never be hyperbolic semitoric, or conversely. 
The main theorem of this paper is the following result about attaching hyperbolic singularities
to an integrable system with non\--degenerate focus\--focus singularities $p_1,\ldots,p_n$, by modifying the system locally near a certain subset $$\{p_i \, |\, i \in S\}$$ of the set of focus\--focus
singularities. By relabeling if needed we may assume that $S=\{1,\ldots,k\}$ for some $k\leq n$.

\begin{theorem} \label{thm:main}
Let $$(M,\Omega,F :=(J,H) \colon M \to \R^2)$$ be an integrable Hamiltonian system
where $J:M\to \R$ is a proper map.
Assume that among the singularities of $F$ are  $m_f$  
simple non\--degenerate focus\--focus singularities
$p_1,\ldots,p_{m_f}$, where $0\leq m_f < \infty$.
Then for any $1 \leq k < m_f$ there exists
an integrable system $$(\widetilde{M},\widetilde \Omega,\widetilde{F}: = (\widetilde  J, \widetilde H) \colon \widetilde M \to \mathbb{R}^2)$$
with the following properties:
\begin{enumerate}
\item[{\rm (1)}] $M = \widetilde M$, $\Omega = \widetilde \Omega$, and $J = \widetilde J$;
\item[{\rm (2)}]
$\widetilde{F}$ has non-degenerate singularities at $p_1,\ldots,p_{k}$ of focus\--focus type; %  $1 \leq i \leq k$
\item[{\rm (3)}]
$\widetilde{F}$  has non-degenerate  singularities at $p_{k+1}, \ldots, p_{m_f}$ of elliptic\--elliptic type;
\item[{\rm (4)}]
There exist $m_f-k$ closed piecewise smooth curves with corners in the classical spectrum $\widetilde{F}(M)$, each of which 
consists of three smooth curves $$\gamma^i_j \colon [0,1] \to \widetilde{F}(M),\,\,\,\,\, j=1,2,3,\,\,\,\,\,\,
k+1 \leq i \leq m_f$$
such that for each fixed $i$ the images
of the interior $(0,1)$ under each of $\gamma^i_1$ and $\gamma^i_2$ consists of non\--degenerate transversally elliptic singular values,
and under $\gamma^i_3$ it consists of non\--degenerate transversally hyperbolic singular values.
The two endpoints where the elliptic and hyperbolic families meet non\---transversally are degenerate 
singularities. The third endpoint is where the two elliptic families meet transversally at the 
non\--degenerate elliptic\--elliptic singularity $p_i$ (see Figure~\ref{fig:Triangle}).
\item[{\rm (5)}]
If $(M,\Omega,F)$ is semitoric (see Definition~\ref{d})
then $(M, \Omega, \widetilde F)$ is hyperbolic semitoric (see Definition~\ref{def:ANDSISH}).
\end{enumerate}
\end{theorem}

We will prove Theorem~\ref{thm:main} in Section~\ref{XX}.

\begin{question} \label{hc}
\emph{Are there are hyperbolic semitoric systems
$$(M,\Omega,F :=(J,H) \colon M \to \R^2)$$
 for which the set of degenerate singularities
is empty?} 
\end{question}

\medskip

The proof of Theorem~\ref{thm:main} in Section~\ref{XX} gives evidence that the answer to the question is
probably going to be \emph{no}, at least under some fairly general conditions. The answer is no in all the physical examples we are aware of.  
That is, it is quite possible that the existence of hyperbolic singularities satisfying (a),(b) and (c.1) 
in Definition~\ref{def:ANDSISH} forces
the existence of at least some degenerate singularities in item (c.2), at the points where the various
non\--degenerate families of singularities under consideration meet with each other. If this were not
the case, there would be loops of singularities, which we cannot rule out at the time being (it would be
crucial to understand whether 
the existence of the global $S^1$\--action may not be compatible with the existence
of such a loop).

\section{The Hopf bifurcation} \label{sec:Hopf}

Another way of formulating the theorem is to say that for a non-degenerate focus\--focus equilibrium point 
in an integrable system it is possible to locally modify the Hamiltonian so that a Hamiltonian Hopf bifurcation 
is induced. The main point is that this modification can be achieved without destroying the integrability of the system. 
Here we are going to present a brief review of some background on the Hopf bifurcation. 
The original references are \cite{MeyerSchmidt71,Sokolskii77,vanderMeer85}.

A Hamiltonian Hopf bifurcation in the linear approximation occurs when pure imaginary 
eigenvalues of an equilibrium collide and move into the complex plane. At the collision point the 
linearization is not diagonalizable. At the bifurcation point the Hamiltonian may be put into 
Sokolskii's normal form 
\[
     \hat H_0 = \omega \Gamma_1 + \sigma \Gamma_2 + C \Gamma_1^2 + 2 B \Gamma_1 \Gamma_3 + 2 D \Gamma_3^2 + h(\Gamma_1, \Gamma_3)
\]
up to flat terms where $h$ contains cubic and higher order terms. 
Here the abbreviations
\begin{align*}
\Gamma_1 &= \hat x  \hat \eta - \hat y \hat \xi  \\
\Gamma_2 &= \frac{\hat \xi^2 + \hat \eta^2}{2} \\
\Gamma_3 &= \frac{\hat x^2 + \hat y^2}{2} 
\end{align*}
are used. Adding the bifurcation parameter $\nu$ gives the unfolding
\[
     \hat H_\nu = \sigma \nu ( a \Gamma_1  + b \Gamma_3) + \hat H_0 \,.
\]
This is the Hamiltonian Hopf bifurcation normal form, where $\omega \not = 0$, $\sigma = \pm 1$, $D \not = 0$, $b \not = 0$.
There are two  different cases, depending on the sign of $\sigma D$. 
The cases $\sigma D < 0$ (respectively $\sigma D > 0$) are called the subcritical (respectively supercritical) Hamiltonian Hopf bifurcation.
When the elliptic\--elliptic equilibrium point looses stability in the subcritical case there is a family of stable
periodic orbits near the unstable focus\--focus equilibrium point. In the supercritical case there is no such family.

Up to flat terms the Hamiltonian Hopf bifurcation normal form $\hat H_\nu$ 
is a family of integrable system 
$$
(\R^4, \Omega, \hat F = (\Gamma_1, \hat H_\nu)),\,\,\,\, \Omega = {\rm d}\hat \xi \wedge {\rm d} 
\hat x  + {\rm d} \hat \eta \wedge {\rm d} \hat y.
$$
To find the image of $\hat F$ and its critical values introduce symplectic polar coordinates 
\begin{align*}
  \hat x &= \sqrt{2 z} \cos\theta,\\
  \hat y &= \sqrt{ 2 z} \sin\theta, \\ 
  2 z p_z &= \hat x \hat \xi + \hat y \hat \eta, \\
  p_\theta &=  \hat x \hat \eta - \hat y \hat \xi 
\end{align*}
so that $$\Omega = p_z \wedge z + p_\theta \wedge \theta.$$
In these coordinates $$\Gamma_1 = p_\theta,$$
$$\Gamma_2 = z p_z^2 + p_\theta^2/(4 z),$$ and
$$\Gamma_3 = z.$$ 
Hence define a reduced Hamiltonian with 
$p_\theta = J$ as a parameter and canonical variables $(z, p_z)$ as
\[
   \hat H_\nu =  \omega J +  \sigma \Big( z p_z^2 + \frac{J^2}{4z} + \nu ( a J + b z)\Big) + C J^2 + 2 B J z + 2  D z^2 \,.
\]
Solving this equation for $p_z(z; J,\hat H)$ gives the reduced action $$\oint p_z(z) {\rm d} z$$ 
and the discriminant of the polynomial 
$$
Q(z) = z^2  p_z(z)^2
$$
 contains the set of critical values of 
$\hat F = (J, \hat H)$. For more details on this derivation using singular reduction 
instead of polar coordinates see \cite{DullinIvanov05}. 
Since $z = \Gamma_3 \ge 0$ only when  the double root is non-negative does 
the corresponding part of the discriminant surface of $Q(z)$ belong to the set of critical values of $\hat F$.

\section{Proof of Theorem~\ref{thm:main}} \label{XX}

We are going to deform the Hamiltonian $H$ to a new Hamiltonian $\widetilde H$ 
in a neighborhood of each focus-focus singularity. The deformation vanishes
outside a sufficiently small neighborhood. We will show that we can choose a deformation
which turns a focus\--focus point into an elliptic\--elliptic point with the properties described in (4). 
This can be done independently for each focus\--focus point, and hence we can create $1 \le m_f - k \le m_f$ 
elliptic\--elliptic points. 

In the construction we use Eliasson's coordinates near each focus\--focus point, 
and we preserve the $S^1$ action which always exist near a focus\--focus point, see Section~\ref{Els}. 
If we are in the semitoric setting, then one of Eliasson's integrals {\em is} the global 
$S^1$ action $J$, and hence the global Hamiltonian $S^1$ action is preserved, establishing  (5).

\medskip

To establish (3) and  (4) for a single equilibrium point there are three steps.

\smallskip
\paragraph{\emph{Step 1}} (A new integrable system) Let $(x, y, \xi, \eta)$ be the Eliasson coordinates given by Theorem~\ref{Eliasson} near $p$ of the original semitoric system and 
let the Hamiltonian of the original system in these coordinates be $H$.
Then there exists a smooth function $G \colon M \to \mathbb{R}$ such that $$\widetilde{H}=H+G,$$ with
$$G=G(J_1,J_2,K_1,K_2),$$ where
\begin{align*}
     J_1 & =  x \eta - y \xi  \\
     J_2 & =  x \xi + y \eta
 \end{align*}
and
\begin{align*}
    K_1 & =  \frac{1}{2}(x^2 + y^2) \\
    K_2 & =  \frac{1}{2}(\xi^2 + \eta^2)
 \end{align*}
Now $J_1, J_2, K_1, K_2$ all have vanishing Poisson bracket with $J = J_1$.
Thus $$\{ \widetilde{H}, J \} = 0$$ and $(M, \Omega, J, \widetilde H)$ is an integrable system.

\smallskip
\paragraph{\emph{Step 2}} (Type of equilibrium)
The quadratic part of the original semitoric integrable system in Eliasson coordinates near the focus\--focus singularity 
is $$H_2 = \omega J_1 + \alpha J_2,$$ with given real parameters $\omega$ and $\alpha$.
% By the non-degeneracy assumption we have $\alpha \not = 0$.
\begin{figure}
\begin{center}
\includegraphics[width=6cm]{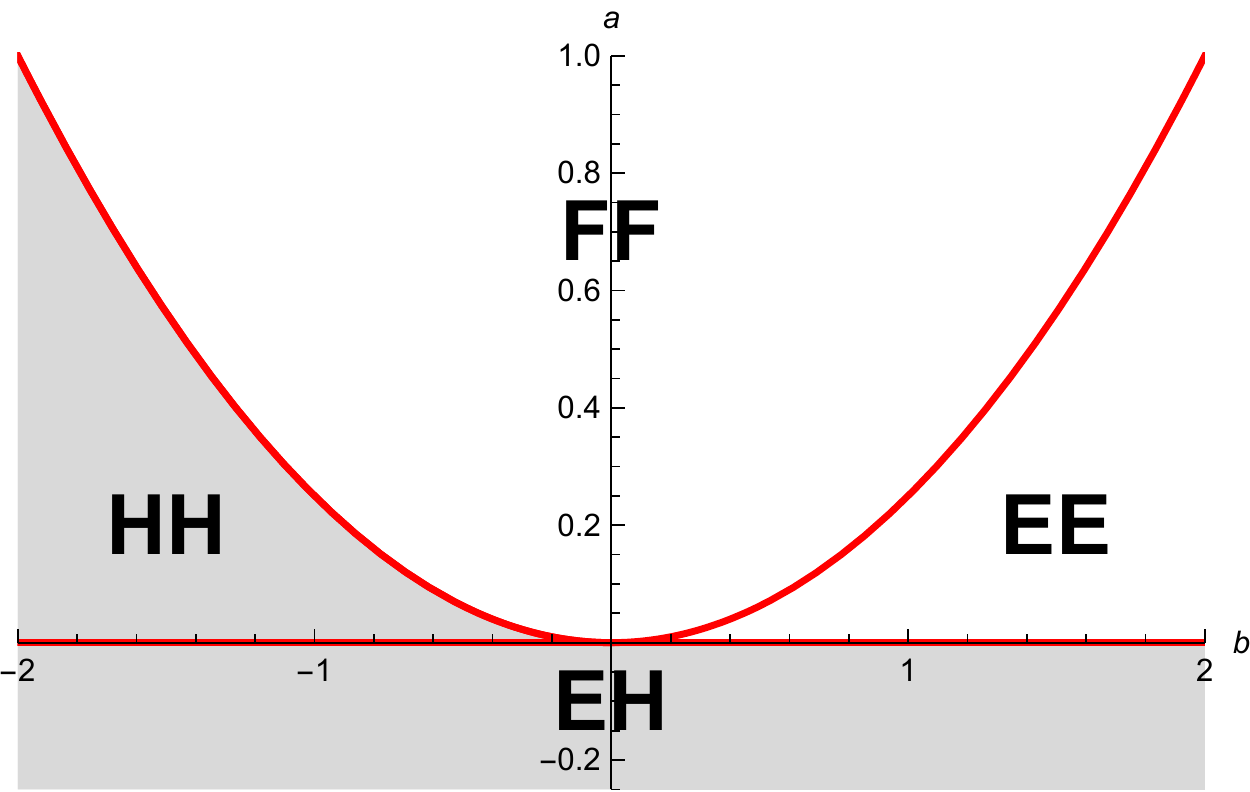}
\end{center}
\caption{Types of roots of the polynomial $P(\lambda) = a + b \lambda^2 + \lambda^4$. 
The discriminant of $P(\lambda)$ is marked in bold (red) lines, delineating the
regions hyperbolic\--hyperbolic (HH) with 4 real roots, elliptic\--elliptic (EE) with 4 pure imaginary roots, elliptic\--hyperbolic (EH) with 
2 real and 2 pure imaginary roots, and focus\--focus (FF) with a complex quadruplet.
For eigenvalues coming from $\widetilde H_2$ the grey region of hyperbolic\--hyperbolic and elliptic\--hyperbolic is not accessible.}
\label{fig:FFEE}
\end{figure}
The quadratic part of the modified integrable system is 
\[
    \widetilde H_2 = \widetilde \omega J_1 + \widetilde \alpha J_2 +  \gamma K_1 + \delta K_2,
\]
with real parameters $\widetilde \omega, \widetilde \alpha, \gamma, \delta$. 
We can choose these  parameters by choosing the function $G$ from Step 1.
We are now going to show that by choosing these parameters 
we can make $\widetilde H_2$ have a non-degenerate elliptic\--elliptic equilibrium point at the origin.
This does not seem to help in creating hyperbolic singularities, but we will see in Step 3 that by adding 
appropriate higher order terms in $G$ we can create hyperbolic\--transverse singularities.

Except for cases with multiple roots the type of equilibrium of $\widetilde H_2$ is determined by the characteristic polynomial
\[
         P(\lambda) = \det ( B \, \Hessien \widetilde H - \lambda) = a + b \lambda^2 + \lambda^4
\]
where $$a = (\widetilde \alpha^2 + \widetilde \omega^2 - \gamma \delta)^2 \ge 0$$ and 
% $$b = -2( \widetilde \alpha^2 -\widetilde \omega^2 - \gamma \delta),$$
$$b = 2( \gamma \delta - \widetilde \alpha^2 +\widetilde \omega^2 ).$$
% see Figure~\ref{fig:FFEE}. 
The discriminant of $P(\lambda)$ is $$16 a ( 4 a  - b^2)^2.$$ 
The line $a = 0$ and the parabola $a = b^2/4$ divide the plane into four regions, see Figure~\ref{fig:FFEE}.
When $ 0 < a < b^2/4$ and $b > 0$ there are four distinct pure imaginary eigenvalues
$$\pm \sqrt{-b/2 \pm \sqrt{ b^2/4 - a}}.$$
Now we verify that by a choice of $\widetilde \alpha, \widetilde \omega, \gamma, \delta$ we 
can satisfy these conditions.
Note that
\[
    b^2/4 - a  = 4 \widetilde\omega^2 (\gamma\delta -  \widetilde\alpha^2)
\]
and by choosing $\gamma \delta > \widetilde \alpha^2$ this can be made positive.
Now $\gamma \delta > \widetilde \alpha^2$ also implies $b > 0$.
Finally we can achieve $a > 0$ by increasing $\gamma \delta$ if necessary.
Since the eigenvalues of the linearization of the Hamiltonian vectorfield of $\tilde H_2$ 
are distinct the singularity at the origin is non-degenerate, 
according to Lemma~\ref{lem:xxx}.
The construction may be done independently at each focus-focus point, 
and thus we established part (3) of the theorem.

\begin{remark} 
Because of the way $a$ and $b$ depend on $\widetilde \alpha, \widetilde \omega, \gamma, \delta$
it is not possible to have $a < 0$ nor is it possible to have $0 < a < b^2/4$ and $b < 0$. 
Thus the idea to directly choose the parameters  $\widetilde \alpha, \widetilde \omega, \gamma, \delta$ so that
$\widetilde H_2$ has a singularity of hyperbolic\--hyperbolic type or 
elliptic\--hyperbolic type does not work.
\end{remark}

\begin{remark}
It is easy to see that when $a = b^2/4$ and $b > 0$ and in addition $\widetilde \omega \not = 0$
the eigenvalues are pure imaginary, the matrix $B \, \Hessien \widetilde H_2$ is not semi-simple,
and the singularity is degenerate.
\end{remark}

\begin{remark}
It is possible to choose a smooth family $\widetilde H_2(s) $ 
with parameter $s \in [0,1]$ such that $\widetilde H_2(0) = H_2$,
$\widetilde H_2(s)$ has a non-degenerate focus-focus singularity for $s \in [0, s^*]$, 
$\widetilde H_2(s)$ has a non-degenerate elliptic-elliptic singularity for $s \in [s^*, 1]$,
and only $\widetilde H_2(s^*)$ is degenerate.
\end{remark}

\smallskip

\begin{figure}
\begin{center}
\includegraphics[width=6cm]{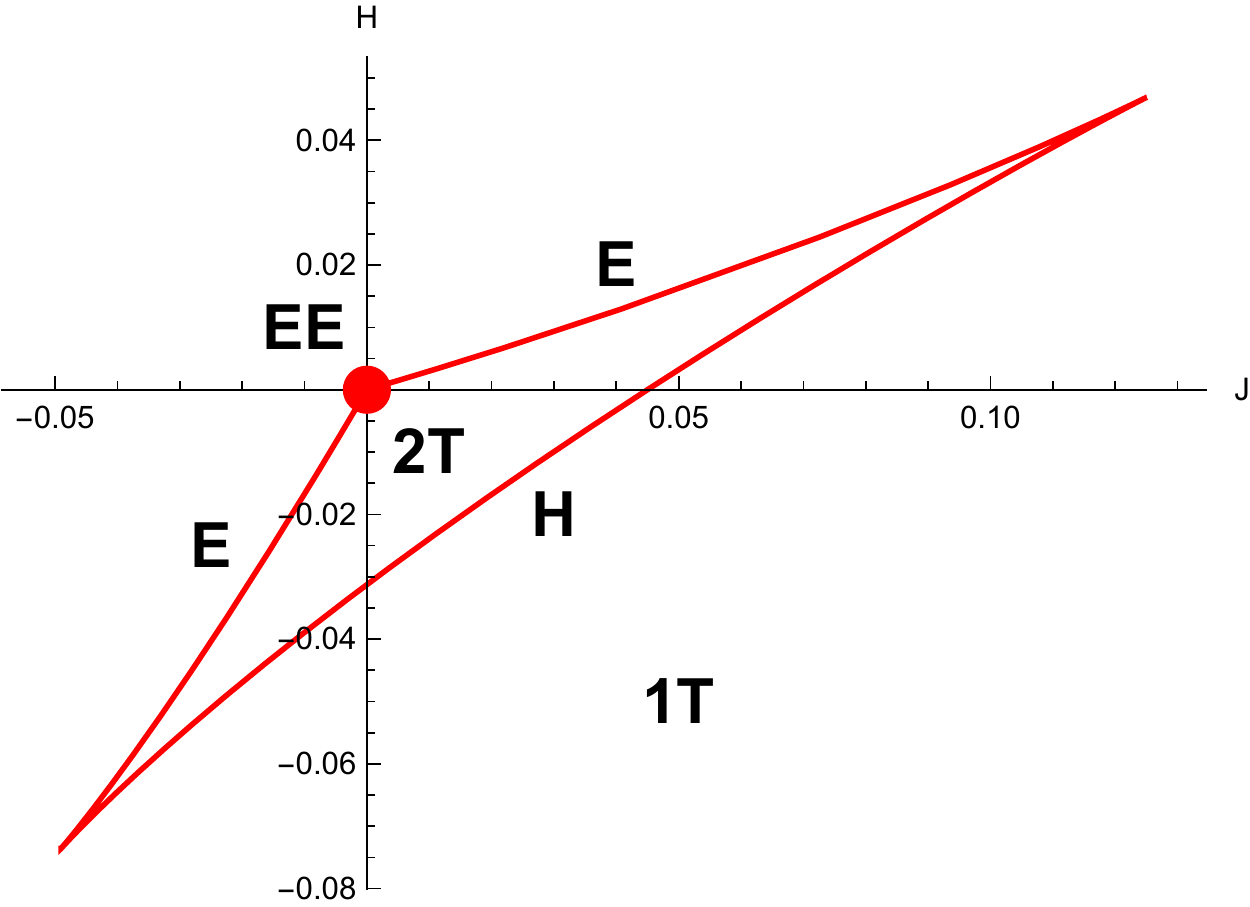}
\end{center}
\caption{Schematic structure of the image of $\hat F$ after the Hopf bifurcation.
elliptic\--elliptic is the non-degenerate elliptic equilibrium point,
E denotes the two transversally elliptic non-degenerate families of singularities (stable isolated periodic orbits) attached 
to the equilibrium,
H denotes the  transversally hyperbolic non-degenerate families of singularities (unstable isolated periodic orbits).
nT denotes that there are n tori in the preimage of the corresponding regular values. 
The cusps where the E and H families meet are degenerate singularities
(saddle-centre bifurcation of periodic orbits). }
\label{fig:Triangle}
\end{figure}

\paragraph{\emph{Step 3}} (Higher order terms) 
We are now going to determine higher order terms in $\widetilde H$ that will generate a family 
of hyperbolic-transverse singularities. This will be done by transforming a special case of the 
general Hopf normal form $\hat H_\nu$ into $\widetilde H$. We are going to show that the image
of $\widetilde F$  has the form schematically shown in Figure~\ref{fig:Triangle} as described 
precisely in the statement part (4) of our main theorem.

To obtain the Hopf normal form we need to introduce a new 
set of local coordinates related to  Eliasson's coordinates by a linear symplectic transformation. 
We now choose a parameter $\hat\gamma$ such that  
\begin{align*}
\widetilde \alpha^2 = \hat\gamma \delta, \,\,\,\,
\widetilde \alpha \not= 0, \,\,\,\,
\widetilde \omega \not= 0, \,\,\,\,
\delta \not = 0,
\end{align*}
and set $\sigma = \sign (\delta).$

The transformation is $$p = T \hat p, \,\, p = (x,y,\xi, \eta),$$ where
\[
T = \begin{pmatrix}
\sqrt{ |\delta| } & 0 & 0  & 0 \\
0 & \sqrt{ |\delta| } & 0 & 0 \\
-\sqrt{ |\hat\gamma |} \sign(\delta \widetilde \alpha)& 0 & 1/\sqrt{| \delta |} & 0 \\
0 & -\sqrt{ |\hat\gamma|} \sign(\delta \widetilde \alpha)& 0 & 1/\sqrt{ |\delta|}  \\
\end{pmatrix} \,.
\]
The transformation $T$ has the property that it maps the quadratic part of $\hat H_\nu$ into $\widetilde H_2$. 
Specifically it transforms the function $\Gamma_i$ as defined in Section~\ref{sec:Hopf} as follows:
\begin{align*}
    \Gamma_1 &= J_1 \circ T, \\
    \Gamma_2 &=\sigma ( \widetilde \alpha J_2  + \hat\gamma K_1 + \delta K_2) \circ T, \\
    \Gamma_3 &=\sigma  (K_1 /  \delta) \circ T \,.
\end{align*}
E.g.\ the first identity for $$J_1 = x \eta -  y \xi $$ follows from 
\begin{eqnarray}
 &
 \sqrt{ | \delta | } \hat x (- \sqrt{ | \hat \gamma|}\, \sign(\hat \gamma \widetilde \alpha)  \hat y + \hat \eta/\sqrt{ | \delta |}  )
- \sqrt{ | \delta | } \hat y ( - \sqrt{ | \hat \gamma|} \,\sign(\hat \gamma \widetilde \alpha)  \hat x + \hat \xi/\sqrt{ | \delta |} ) 
 \nonumber \\
& =  \hat x \hat \eta  - \hat y \hat \xi  \nonumber
\end{eqnarray}

We now specialize the Hopf normal form $\hat H_\nu$ discussed in 
Section~\ref{sec:Hopf} to $$a = B = C = 0, \,\,\,\,b = 1,$$ because that is sufficient for our purpose.
Transforming the Hopf normal form (given in variables with hat) and with sufficiently small $\nu$
\[
     \hat H_\nu = \widetilde \omega \Gamma_1 + \sigma ( \Gamma_2 + \nu \Gamma_3)  + 2 D \Gamma_3^2 % + C \Gamma_1^2 + 2 B \Gamma_1 \Gamma_3
\]
into variables without hat gives
\[
    \widetilde H = \widetilde \omega J_1 + \widetilde \alpha J_2 + \left( \hat \gamma + \frac{\nu}{\delta}\right) K_1 + \delta K_2 
     + 2  D ( \alpha J_2 + \hat \gamma K_1 + \delta K_2)^2 \,. % + C J_1^2 + 2 \frac{B}{|\delta|} J_1 K_1
\]
Thus the quadratic part of $\widetilde H$ has parameters in the notation of Step 2 given by
$\widetilde \omega, \widetilde \alpha, \delta$ and $$\gamma = \hat\gamma + \nu/\delta.$$
Now we choose the function $G$ by setting $$G = \widetilde H - H,$$ 
so that $$H + G = \widetilde H.$$ Note that $G$ is not only producing the desired $\widetilde H$, 
but it also annihilates any unwanted higher order terms which may be present in $H$.

Recall that $$\widetilde \alpha^2 = \hat\gamma \delta, \,\,\widetilde \alpha \not= 0,$$ and $\widetilde \omega \not= 0$,
and therefore the parameters in Step 2 are $$\widetilde \omega, \widetilde \alpha, \gamma = \hat \gamma + \nu/\delta, \delta,$$ 
and they put $\widetilde H$ in the elliptic\--elliptic region as long as $\nu$ is positive. 
This may be seen by computing the eigenvalues of the linearized vector field of $\hat H_\nu$ at the origin:
\begin{itemize}
\item
when $\nu < 0$ the equilibrium is of type focus\--focus with eigenvalues $$\pm \sqrt{ -\nu} \pm {\rm i} \widetilde \omega$$
\item 
when $\nu > 0$ it is of type elliptic\--elliptic with eigenvalues $$  \pm {\rm i} ( \sqrt{\nu} \pm \widetilde \omega).$$
\end{itemize}
% Alternatively we can specialise the formulas from step 2.
Hence for sufficiently small and positive $\nu$ the equilibrium is of 
elliptic\--elliptic type and it is non-degenerate because the eigenvalues are distinct.

\begin{figure}
\begin{center}
\includegraphics[width=9cm]{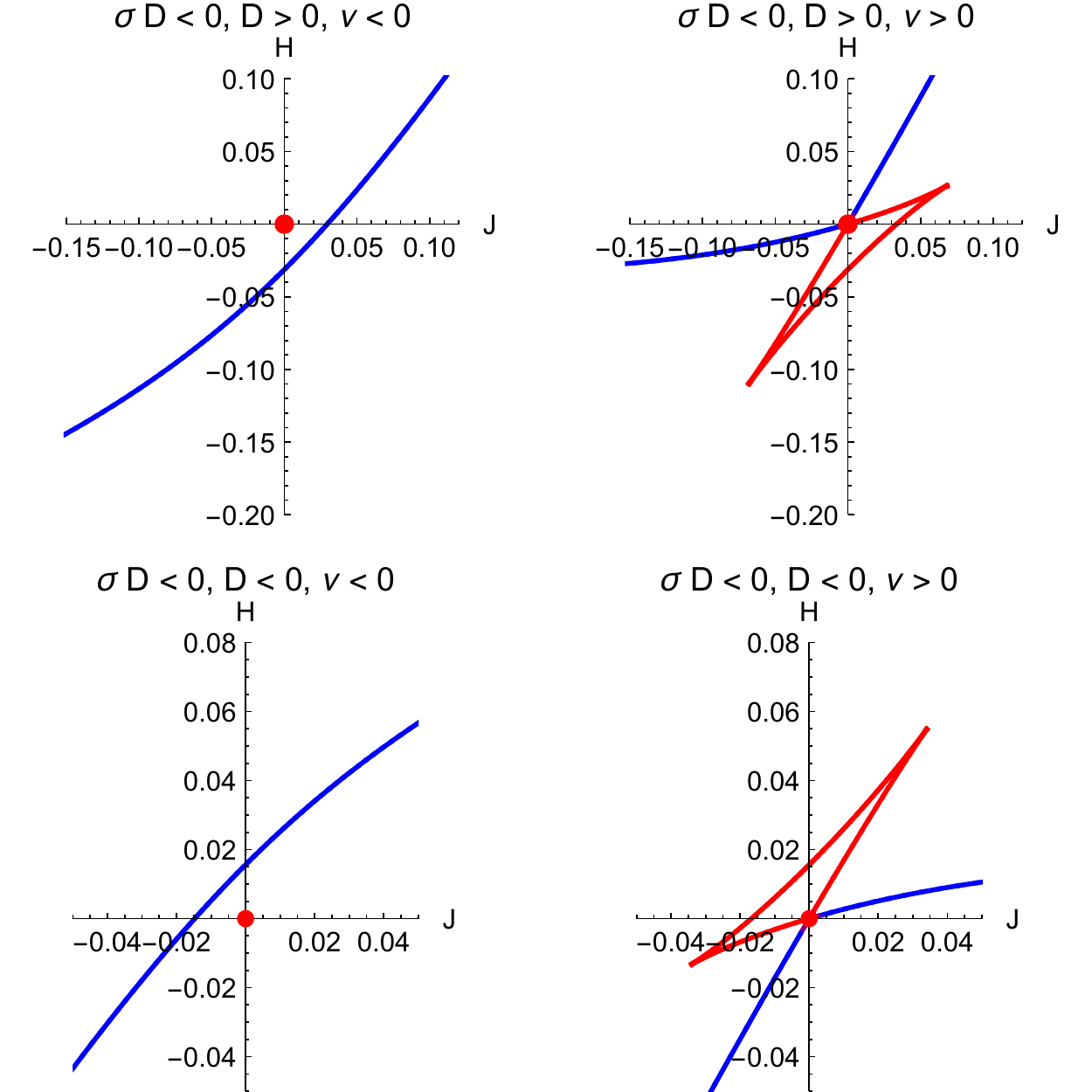}
\end{center}
\caption{Image of the map $\hat F$ for the subcritical Hopf bifurcation with $\sigma D < 0$.
Critical values of $\hat F$ are shown in red. In addition, the vanishing  of the discriminant of $Q(z)$ is shown in blue. 
The parameter values are $\omega = 1$, $\nu = \pm 1/2$, and $D = 1$ or $D = -2$.}
\label{fig:Hopf1}
\end{figure}

\begin{figure}
\begin{center}
\includegraphics[width=9cm]{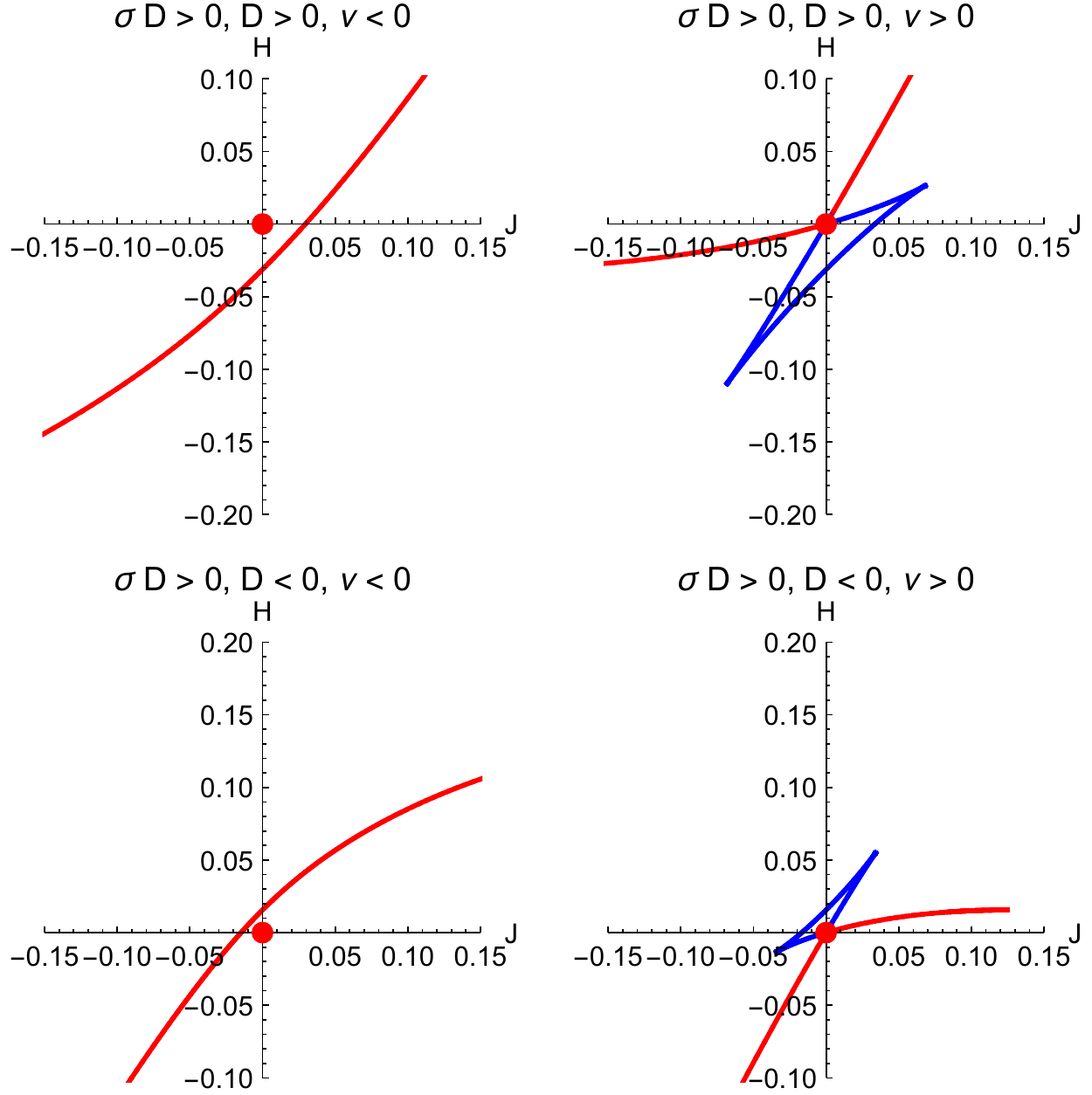}
\end{center}
\caption{Image of the map $\hat F$ for the supercritical Hopf bifurcation with $\sigma D > 0$.
Critical values of $\hat F$ are shown in red. In addition, the vanishing  of the discriminant of $Q(z)$ is shown in blue. 
The image is to the left for $D > 0$ and to the right for $D < 0$. 
The parameter values are $\omega = 1$, $\nu = \pm 1/2$, and $D = 1$ or $D = -2$.}
\label{fig:Hopf2}
\end{figure}

We are now going to establish the properties of the image of $\widetilde F$ claimed in (4) of the main theorem. 
See Figure~\ref{fig:Hopf1} and \ref{fig:Hopf2} for an illustration of the various cases. 

First of all notice that $\widetilde F$ and $\hat F$ are related by a linear symplectic transformation,
so we may as well study the image of $\hat F$.
As described in Section~\ref{sec:Hopf} the critical values of $(\Gamma_1, \hat H_\nu)$ are contained 
in the discriminant surface of the cubic polynomial 
\[
    Q(z) = 4 z \sigma (\hat H - \omega J - \sigma \nu z - 2 D z^2)  - J^2 \,. %  - C J^2 - 2 B J z
\]
The discriminant of this polynomial has a rational parametrization that can be found by 
the Ansatz 
\[
Q(z) = - 4 \sigma  D (z - d)^2 ( 2 z + \sigma s^2 / D ) 
\]
and solving for the double root $d$, $J$ and $\hat H$. The result is 
\[
%   (J_c(s), \hat H_c(s)) = \left( \sigma s \frac{ s^2 - \nu}{2(D + B \sigma s)},  \omega J_c(s) + C J_c(s)^2 + s J_c(s) -D \frac{J_c(s)^2}{2s^2}  \right) \,,
%   (J_c(s), \hat H_c(s)) = \left( \frac{ 1}{2D} s ( s^2 - \nu),  \omega J_c(s) + s J_c(s) -D \frac{J_c(s)^2}{2s^2}  \right) \,,
   (J_c(s), \hat H_c(s)) = \left( \frac{ 1}{2D} s ( s^2 - \nu),  \frac{1}{4s} J_c(s) ( \nu + 4 s \omega  + 3 s^2)  \right) \,,
\]
where the double roots of $Q(z)$ occur at $$z = \Gamma_3 = d(s) = \sigma J_c(s)/( 2 s).$$

Assume that $\nu > 0$.  Three piecewise smooth curves of critical values are found
for $s \in [-\sqrt{\nu}, - \sqrt{ \nu/3}]$, 
for $s \in [ -\sqrt{\nu/3}, \sqrt{ \nu/3}]$,
and $s \in [ \sqrt{ \nu/3}, \sqrt{ \nu}]$.
Both derivatives 
\[
    \frac{{\rm d}}{{\rm d}s} (J_c(s), \hat H_c(s) )  = \frac{3 s^2 - \nu}{2D} (1, s + \omega ) \,.
\]
vanishes at $s = \pm \sqrt{ \nu/3}$ only. We call these corresponding critical values  the cusp. 
The other segment endpoints $\pm \sqrt{ \nu}$ map to the same critical value $(0, 0)$.

We now establish the type and the (non-)degeneracy of the critical values 
on the three curves.
We need to compute the determinant of the Hessian of the reduced Hamiltonian $\hat H(z, p_z)$ from section~\ref{sec:Hopf}
at the critical points. %  for $s \in (-\sqrt{\nu}, \sqrt{\nu} )$. 
The determinant of the Hessian at an arbitrary point is $$-4 p_z^2 + 8 \sigma D z + J^2/z^2,$$
and evaluating this at $J = J_c(s)$, $z = d(s)$, and $p_z= 0$ gives $$2( 3 s^2 - \nu),$$ which is non-zero 
unless $s = \pm \sqrt{ \nu/3}$, corresponding to the cusps. Moreover for 
$s \in ( -\sqrt{ \nu/3 } , \sqrt{ \nu/3})$ the determinant is negative, so this segment is transversally\--hyperbolic.
Similarly, the other two segments $ (- \sqrt{ \nu}, - \sqrt{ \nu/3}) $ and  $ ( \sqrt{ \nu/3},  \sqrt{ \nu}) $
lead to a positive determinant and are hence transversally\--elliptic.
Finally at the cusps the determinant of the Hessian vanishes and hence these are degenerate critical values.

The three curves are graphs over $J$ because $$\frac{{\rm d}J_c(s)}{{\rm d}s} \not = 0$$ on the interior of the three segments,
and when $$\frac{{\rm d}J_c(s)}{{\rm d}s} = 0$$ then also $$\frac{{\rm d}\hat H_c(s)}{{\rm d}s} = 0,$$ creating the two cusps.
When the two curves of transversally elliptic values intersect at the origin they have slopes 
$$\omega \pm \sigma \sqrt{ \nu},$$ 
and hence the intersection is transversal for $\nu > 0$.

Not all points in the discriminant surface of $Q$ are actually critical values of $\hat F = (J, \hat H)$. 
To be in the image of $\hat F$ we need $\Gamma_3 \ge 0$. The double 
root $$d(s) = \sigma (s^2 - \nu)/(4 D)$$ is non-negative for $s \in [-\sqrt{\nu}, \sqrt{\nu}]$ only when 
$\sigma D < 0$. Hence we are now choosing $D$ such that $\sigma D < 0$. 
Accordingly we have one of the two right situations shown in Figure~\ref{fig:Hopf1}.

A particular point on the transversally\--hyperbolic curve
is obtained from $s=0$ where $$(J, \hat H) = ( 0 , -\nu^2/ ( 8 D)).$$
Hence the sign of $D$ determines whether this critical value is above or below the equilibrium point at the origin.

\begin{remark} 
The cusps corresponding to $s = \pm \sqrt{ \nu/3}$ are created by a saddle-centre bifurcation of an elliptic and a hyperbolic periodic orbit.
This is one of the typical degenerate singularities described in \cite{BolsinovFomenko04}.
For a detailed analysis of the universal features of the dynamics near this bifurcation using an integrable model see \cite{DullinIvanov05b}.

\end{remark}

\begin{remark}
As in Step 2, we can think of the function $G$ as a one-parameter family that smoothly 
deforms the original Hamiltonian into the new Hamiltonian $\widetilde H$. From this 
point of view it makes sense to consider $\nu < 0$ as well, and the corresponding 
sets of critical values of $\hat F$ are shown on the left column of Figure~\ref{fig:Hopf1} and \ref{fig:Hopf2}.
\end{remark}

\section{Extensions}

The main result of this paper is stated in the form of several theorems, but in fact it is
more a method than a result. The following Theorem has a similar proof.

\begin{theorem} \label{thm:noFF}
Each focus\--focus singular point in a semitoric integrable system $(M, \Omega, F)$ may be  continuously deformed
via a path of semitoric systems into an elliptic-elliptic point on the boundary of the image of the new integrable system $(M,\Omega,\widetilde F)$ 
constructed in Section~\ref{XX}, either at the top boundary 
(maximal $\widetilde H$ for fixed $J$)
or at the bottom boundary (minimal $\widetilde H$ for fixed $J$). 
This may be done simultaneously for any subset of the set of focus-focus singular points of the system.
\end{theorem} 
The proof is as above, the only difference is that now we choose $G$
with $\sigma D > 0$. The corresponding critical values are shown 
in the two right cases in Figure~\ref{fig:Hopf2}.
The sign of $D$ determines which one of the two cases occurs.

Under what conditions the resulting semitoric system is in fact toric is an open question.

\section{Examples} \label{sec:Examples}

A deformation of the spherical pendulum that models floppy triatomic molecules (eg.\ HCN) has been studied in 
\cite{Efstathiou05}. Strictly speaking this is not an example for out theorem since the momentum 
map $J$ of the spherical pendulum is not proper. But if $J$ is not proper the same idea still works
as long as $\hat F$ is proper, which can be easily checked in this particular example. In fact, 
our theorem can be generalised to the situation where $J$ is not proper by adding higher order terms
to $G$ that make $\widetilde H$ proper.

\begin{figure}
\begin{center}
\includegraphics[width=7cm]{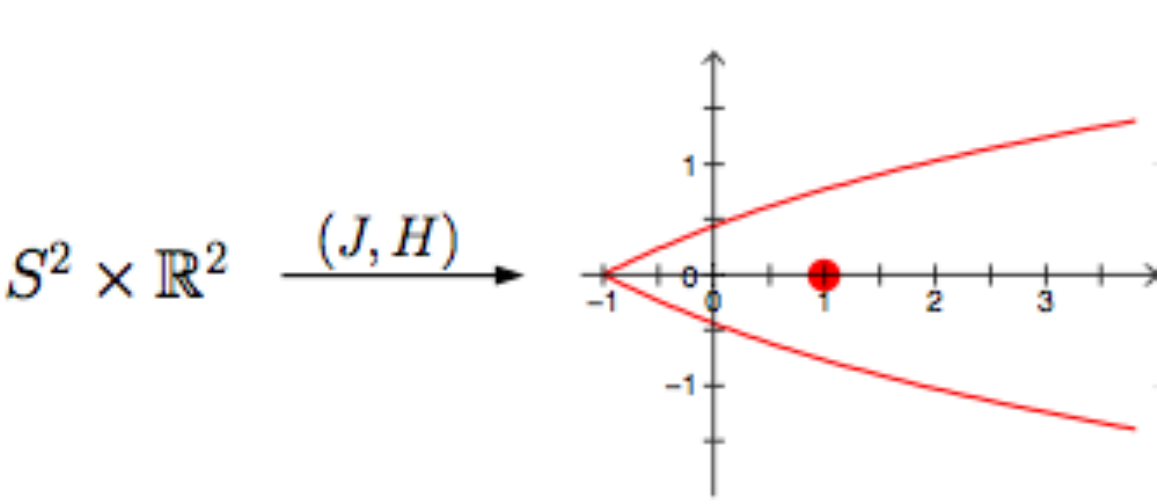}
\end{center}
\caption{The image of the of the Jaynes\--Cummings system $(J,H)\colon S^2\times\R^2 \to \R^2$ is bounded by
the  curve in the figure which has a cusp at $(-1,0)$. The dot corresponds to the image of the only focus\--focus
singularity of the system.}
\label{fig:JC}
\end{figure}

Here we give a brief description of a deformation of the semitoric coupled spin-oscillator 
\cite{PeVN12} (Jaynes\--Cummings system)
that makes it pass through a subcritical Hopf bifurcation with $\sigma D > 0$.
A family for which the focus\--focus point is driven into the boundary in a supercritical Hopf bifurcation with $\sigma D < 0$  
has been presented in \cite{VN07}. The image of the momentum map of the Jaynes-Cummings
is shown in Figure~\ref{fig:JC}

In particular examples it may be simpler to avoid the Eliasson normal form, since it may not be easy to find. 
Instead we start with $F = (J, H)$ in the original variables $(u, v, x, y, z)$ 
where $$H = ( x u + y v)/2$$ and $$J = (u^2 + v^2)/2 + z.$$
Then we ask for a function 
$G$ that commutes with $J$, and  define $$\hat H = H + G.$$
In an example it will also typically be nicer to work with a global function $G$, instead of one 
that is only defined in a neighborhood of the equilibrium under consideration.

For the spin-oscillator any function of the form $$G( z, v x - u y, u^2 + v^2, x u + v y)$$ works.
We chose $G = G(z)$, and the characteristic polynomial of the linearization at the equilibrium point $(0,0,0,0,1)$
is given by $$\frac{1}{16} - \frac12 \lambda^2(1 - 2 G'(1)^2 ) + \lambda^4$$
and undergoes a Hopf bifurcation when $G'(1)^2 = 1$. Now set $$G(z) = \gamma z^2.$$
In the $(b,a)$ plane we obtain the horizontal line $$(b,a) = (1/2 - 8 \gamma^2, 1/16).$$
For $\gamma > 1/2$ the singularity is in the elliptic\--elliptic region.
The non-linear analysis of this example is somewhat involved and will be presented in detail 
in a forthcoming paper. Here we simply present the image of $\hat F$ for 
$\gamma = 4/5$, see Figure~\ref{fig:spinoscHopf}.

\noindent
\medskip\noindent

\smallskip\noindent
 Holger R.~Dullin\\
School of Mathematics and Statistics\\
University of Sydney\\
Sydney, NSW 2006, Australia\\
{\em E\--mail}: \texttt{holger.dullin@sydney.edu.au}
\noindent
\\
\\
{\'A}lvaro Pelayo \\
University of California, San Diego\\ 
Mathematics Department \\
9500 Gilman Dr \#0112\\
La Jolla, CA 92093-0112, USA.\\
{\em E\--mail}: \texttt{alpelayo@math.ucsd.edu}

\end{document}